\documentclass[ars, manuscript]{copernicus}
\usepackage{graphicx} 
\begin{document}
\nolinenumbers 
\title{Experiments with an oscillator based Ising machine}
\Author[1,*][roy@anabrid.com]{Shrish}{Roy} 
\Author[2,*]{Bernd}{Ulmann}
\affil[1]{anabrid GmbH, Franklinstra\"se 56, 60486 Frankfurt/Main, Germany}
\affil[2]{anabrid GmbH; FOM University of Applied Science, 60486 Frankfurt/Main, Germany}
\affil[*]{These authors contributed equally to this work.}

\runningtitle{TEXT}     
\runningauthor{TEXT}    

\received{}
\pubdiscuss{} 
\revised{}
\accepted{}
\published{}

\firstpage{1}

\maketitle
\begin{abstract}
Interest in non-algorithmic, unconventional computing is rising in recent years due to more and more apparent short comings of classic stored-program digital computers, such as energy efficiency, degree of parallelism in computations, clock frequency limitations, integration density, silicon utilization, etc. One notable such unconventional approach are \emph{oscillator based \textsc{Ising} machines}, i.\,e., systems consisting of a number of oscillators which can be coupled in order to create an analogue for some problem to be solved, while the actual information is encoded in the phase relationships of these oscillators with respect to some reference (typically one of these oscillators). It has been shown that machines of this type are capable of solving NP-hard problems such as max-cut, etc. In the following an experimental \textsc{Ising} machine is presented together with experimental results obtained from this machine.
\end{abstract}
\introduction  
There is a number of reasons why to embark on the journey described in the following: Imagine yourself grappling with an intricate puzzle, where the mission is to uncover the optimal arrangement of pieces that seamlessly align with specified conditions. Typical examples are schedule optimization, protein folding, solving intricate problems on graphs, etc. Oscillator based \textsc{Ising} machines as described here show promise as cutting-edge tools to solve such puzzles and problems, more generally referred to as optimization problems, which are ubiquitous not just in an engineering context but many other disciplines. These machines harness a concept derived from physics, specifically a model initially employed to decipher the properties of magnetism. In this \textsc{Ising} paradigm, due to \textsc{Ernst Ising}, 1925 (\cite{ising}), atoms (more precisely their spins) are reduced to being either \emph{up} or \emph{down}, akin to a binary switch, and the machine's exhilarating task is to identify an optimal combination of these states, satisfying certain rules and requirements. From an engineer's perspective, each up or down might represent a critical state decision in a highly complicated system or the like. Such machines excel at certain optimization problems and can outperform even the fastest classic digital computers without having to resort to complicated quantum computers.

Interestingly, these \textsc{Ising} machines exhibit traits normally found in \emph{adiabatic quantum computers} at a fraction of the latter's complexity and price, bestowing upon them potential capabilities that reach beyond what traditional computers can achieve, all without mandating fully quantum systems. This feature is particularly attractive when facing the daunting realm of extraordinarily challenging large-scale optimization problems. The adoption of \textsc{Ising} machines could result in groundbreaking improvements in efficiency in domains such as network design, resource allocation, and system optimization; areas where optimal solutions are of utmost importance. Such machines are easily interfaced with classic stored-program digital computers and may act as specialized co-processors for certain highly compute intensive tasks.

The machine(s) described in the following were built using off-the-shelf analog computer components together with simple phase-shift-oscillators. No special parts hard to procure were used, making this system suitable as a conceptual base for similar implementations and well suited not just for experimentation but also for student work.
\section{\textsc{Ising} machines}
Computational science and mathematics provide distinct frameworks for classifying the complexity of problems. Among these is the classification of \emph{NP-hard} problems, which are characterized by the computationally intensive nature of solution verification. Solving an NP-hard problem requires significant computational resources if it is even feasible at all. Such problems appear in a wide range of disciplines, including physics, biology, engineering, as well as even recreational games. The general \textsc{Ising} model is mathematically represented as follows (\cite{wang2019oim}):
\begin{equation}\label{eq:ising_model_equation}
    H=-\sum_{i,j}J_{ij}\sigma_i \sigma_j-\sum_j h\sigma_j,
\end{equation}

This equation can be conceptualized as a magnetic structure or material composed of tiny particles exhibiting spins $\sigma_i\in\{-1,+1\}$ (indicating a downward or upward spin). The coupling coefficient $J_{ij}\in \mathbb{R}$ signifies a spin-spin interaction term between neighboring particles and an external field $h$. (The equation is often simplified in the literature by omitting the second summation term.) The sum of these values results in a Hamiltonian $H$ that delineates the energy of the whole system. 

The \textsc{Ising} problem now seeks a spin arrangement of these particles (given a magnetic structure with a specified interaction term) yielding a minimal energy configuration for the entire system. Consequently, \ref{eq:ising_model_equation} becomes an optimization problem. It is recognized that NP problems (decision problems solvable by a nondeterministic machine in polynomial time) can be transformed into NP-hard problems utilizing polynomial resources. 

An oscillator based \textsc{Ising} machine as described in the following can now be used to solve such problems by configuring it in accordance to the Hamiltonian describing the problem under consideration. A configuration in this particular context is a graph showing the connections between oscillators with adjustable connection weights. The machine then progresses towards the Hamiltonian's lowest-energy state by letting the oscillators synchronize with eachother (\cite{PIK01}), i.\,e., it converges towards the ground state, thus yielding the desired optimum solution (\cite{mohseni2022ising}, \cite{bian2010ising}, \cite{caltech}). 

These coupled oscillators may achieve synchronization in phase or out-of-phase (antiphase) with respect to some ``master oscillator''. The system described here uses antiphase synchronization; that is, two mutually coupled oscillators will synchronize so that one oscillator is out of phase with the oscillator it is interacting with by $\pi$. Consequently, \emph{maximum cut} problems (\emph{max-cut}) are an optimal problem class for this type of architecture.
\section{$NP$ and $NP$-hard Problems}
Complexity classes form the backbone of computational complexity theory, providing a framework to categorize problems based on their inherent difficulty. Of these, three key complexity classes stand out: $P$, $NP$, and $NP$-\emph{hard}. These classes offer crucial insights into algorithmic efficiency and problem solvability. 

Class $P$ (\emph{polynomial time}) contains problems that can be solved efficiently by deterministic machines such as classic stores-program digital computers. Formally, a problem belongs to P if there exists a deterministic Turing machine $M$ and a polynomial $p(n)$ such that $M$ halts in $\mathcal{O}(p(n))$ time for every input $x$. This class represents problems that are considered \emph{tractable} and feasible to solve in practice. \textsc{Dijkstra}'s algorithm for finding the shortest path in a graph is a prime example of a solution for a problem in $P$. The algorithm runs in polynomial time relative to the number of vertices and edges in the graph, making it efficient and practical for real-world applications.

NP (\emph{nondeterministic polynomial time}) contains problems whose solutions, once proposed, can be verified quickly (in polynomial time) by a deterministic algorithm. Mathematically, a problem is in $NP$ if it can be solved by a \emph{nondeterministic} Turing machine $M$ in polynomial time $p(n)$. A classic example for a problem in this class is the \emph{boolean satisfiability problem} (\emph{SAT}). Finding an actual solution for this problem is very challenging, while its verification it computationally efficient.

$NP$-hard problems are at least as difficult as the hardest problems in $NP$. A problem is $NP$-hard if every other problem in $NP$ can be reduced to it in polynomial time. This means that if one $NP$-hard problem could be solved efficiently, all problems it can be transformed into can be solved as well. To demonstrate the $NP$-Hardness of SAT, one could perform a polynomial-time reduction from the \emph{Hamiltonian cycle problem} (\emph{HCP}) to SAT. This reduction would transform any instance of HCP into a corresponding SAT instance in polynomial time.

Understanding these complexity classes is crucial for algorithm design and system optimization in general. Problems in $P$ are generally considered solvable in practice, while $NP$ and $NP$-hard problems often require heuristic approaches or approximation algorithms for large problem instances. The relationship between these classes, particularly the question of whether $P=NP$, remains one of the most significant open problems in theoretical computer science. If $P=NP$, it would revolutionize our approach to many computational problems, potentially leading to efficient solutions to currently intractable problems in areas such as cryptography, optimization, and artificial intelligence. As the boundaries of computational efficiency are pushed forward, a deep understanding of these complexity classes provides valuable insights into the fundamental limits and possibilities of algorithmic solutions to complex problems (\cite{erementchouk2022computational}).
\section{Synchronization}
The synchronization of oscillators is a rather intricate concept at first glance. However, it is a phenomenon frequently encountered in daily life. An oscillator in general is any entity that exhibits regular motion or cycles, such as the pendulum or the balance wheel in a mechanical clock, a vibrating  string, etc. These oscillators adhere to a defined rhythm or cycle, which they perpetually repeat. Synchronization involves aligning these oscillators so that their frequencies align with a certain phase relationship. 

The synchronization of oscillators requires some coupling mechanism so that some influence may be established between these oscillators. A classic example consists of two or more metronomes standing on a common platform that is free to move (slightly) in the direction the metronome's pendulums swing in. Starting with metronomes set to similar resonant frequencies the will synchronize over time all not only locking at a common frequency but all oscillating in the same phase. Synchronization effects are found in nearly all physical and natural systems, sometimes desirable, sometimes undesirable (\cite{strogatz2018nonlinear}). In the context of oscillator based \textsc{Ising} machines, the synchronization of a plethora of oscillators is at the heart of the system.

Oscillators synchronized with a phase difference of $\pi$ are called to be in \emph{antiphase}, if the phase difference is $0$, they are \emph{in phase}. In practice these values are only closely approximated but never perfectly met. It should also be noted that inadvertent coupling between oscillators can become a challenge as even tiny coupling coefficients due to shared power rails, etc., may result in such a coupling. It is therefore necessary to decouple the power rails of the individual oscillators as good as possible.
See \cite{yuan2017multistable} for details on tuning \textsc{Ising} machines.

Typically the phase differences will themselves oscillate around $0$ or $\pi$. Understanding these dynamics is essential for interpreting transient behaviors as the machine seeks optimal states (\cite{kawamura2014phase}).
The specific stable state achieved for a certain coupling topology depends on initial conditions and operational parameters (\cite{yuan2017multistable}).
Factors such as coupling strength and delay also influence the emergence of specific modes, critical for designing \textsc{Ising} machine interaction networks (\cite{strogatz2018nonlinear}).

For systems with multiple oscillator groups, a $\pi$ phase difference between collective rhythms can lead to effective antiphase synchronization at a macroscopic level, even when individual oscillators are coupled in phase at a microscopic level (\cite{yuan2017multistable}).

The role of a $\pi$ phase difference is pivotal for computational strategies of \textsc{Ising} machines. In contrast to antiphase synchronization, in-phase synchronization emerges as an equally critical phenomenon, characterized by the alignment of all (or at least many) oscillators within a system, causing them to oscillate in unison at the same frequency. This synchronization mode optimizes coherence, thereby enhancing the overall efficiency of the system and increasing its resilience to disturbances. In-phase synchronization is particularly vital in disciplines such as cardiac physiology and electrical engineering, where it ensures the constructive combination of outputs, resulting in stronger, more cohesive signals and more effective operations (\cite{strogatz2018nonlinear}).

\section{Mathematics of synchronization}
The conceptual mathematical framework used to study the phenomenon of synchronization in complex systems of coupled oscillators is the \textsc{Kuramoto} model. Developed by \textsc{Yoshiki Kuramoto} in the 1970s, it provides profound insights into how large groups of oscillators, such as cells, neurons, or oscillators in the case of machines as described here, achieve synchronization spontaneously through simple interactions (\cite{PIK01}, \cite{strogatz2018nonlinear}).

The \textsc{Kuramoto} model considers a collection of $N\in\mathbb{N}$ oscillators, each with its own natural frequency, which are coupled together so that their phases influence each other. The basic form of the model is expressed by the following differential equation for each oscillator $1\leq i\leq N$:

\begin{equation}
\dot{\theta}_i = \omega_i + \frac{K}{N} \sum_{j=1}^N \sin(\theta_j - \theta_i)    
\end{equation}

Here, $\theta_i$ represents the phase of the $i$-th oscillator, $\omega_i$ is its natural frequency drawn from a given frequency distribution, and $K$ is the coupling strength (a measure of how much oscillators influence one another). The sine term represents the phase difference between oscillators, eventually driving them towards synchronization.

To measure the degree of synchronization in the system, the \textsc{Kuramoto} model uses an order parameter $r$, defined as
\begin{equation}
r e^{i\psi} = \frac{1}{N} \sum_{j=1}^N e^{i\theta_j}.
\end{equation}

In this expression, $0<r<1$ is the magnitude of the order parameter. When $r$ is close to 0, the oscillators are out of sync and their phases spread out over time. $r$ being close to 1 indicates that a significant fraction of the oscillators are synchronized, sharing a similar phase. The variable $\psi$ represents the average phase of the population.

The beauty of the \textsc{Kuramoto} model lies in its inherent simplicity and applicability to a wide range of real-world systems. It shows how synchronization can emerge even from simple interactions and how this synchronization critically depends on the coupling strength and the diversity of natural frequencies within the oscillator population. This model has been pivotal in the advancement of understanding complex dynamical systems in various scientific disciplines, including physics, biology, and engineering. Since the machine described in this paper is still in its primitive stages, its parameters are yet to be determined and detailed analysis of its synchronization behavior is still pending.
\section{Max-cut Problems}
\begin{quote}
  Given an undirected edge-weighted graph $G(V,E)$, the maximum cut problem (Max-Cut) is to find a bipartition of the vertices that maximizes the weight of the edges crossing the partition.  
\end{quote}

In this context, \(V\) denotes the number of vertices and \(E\) represents the number of edges in the graph. In the Max-Cut problem, the objective is to divide the vertices into two distinct groups so that the count of edges that span these groups is maximized. This problem can be represented by the \textsc{Ising} model, where each vertex is associated with a spin variable that may assume one of two possible values. In the context of the \textsc{Ising} model, positive (antiferromagnetic) couplings between spins induce neighboring spins to take opposite orientations. When two adjacent spins are oriented oppositely, the respective edge is "cut" by the partition, resulting in a reduction in the system's energy. Consequently, minimizing the energy in the \textsc{Ising} model, by identifying a spin configuration that yields low energy, effectively maximizes the number of edges cut in the original Max-Cut problem.

The max-cut problem is a prominent challenge in computer science and mathematics, particularly within optimization and graph theory. A graph comprises nodes interconnected by edges, and the objective in the context of the max-cut problem is to segregate these nodes into two separate groups. The primary aim here is to maximize the number of edges (or the sum of the edge weights) that have endpoints within both groups. Essentially, the goal is to have as many edges as possible traversing between the two groups rather than residing within the same group. 

Solutions of max-cut problems is required in various scientific fields. For example, in network design, it is critical to ensure robust communication between two subnetworks. In circuit design and layout, the division of components into two segments can help minimize interconnection costs or optimize specific performance metrics. 

The max-cut problem is characterized as $NP$-hard, indicating the absence of a known efficient algorithm to determine an exact solution for all generalized cases as the problem size increases. Nonetheless, various approximation algorithms and heuristics have been developed to derive satisfactory solutions within a reasonable timeframe in practical applications. This problem serves as a fundamental example in computer science to illustrate the complexities of optimization and the methodologies adopted to address computationally intensive tasks.
\section{Suitability for anti-phase synchronized hardware}
The max-cut problem is well-suited for an antiphase synchronization model due to the inherent alignment between the problem's structure and the dynamics of the oscillators and their synchronization scheme. In the max-cut problem, vertices are partitioned into two distinct sets, mirroring the binary characteristic of antiphase synchronization where oscillators are confined to one of two possible states, specifically the $0$ or $\pi$. The objective of maximizing the cut size in the max-cut problem aligns with the minimization of energy in the coupled oscillator system. Edges that connect vertices situated in different sets, thus contributing to the cut, are represented by oscillators in opposing phases, which constitutes the most energy-efficient state in anti-phase synchronization. The edges within the graph are depicted by the coupling interactions between oscillators. In the context of anti-phase synchronization: 
\begin{itemize}
    \item Oscillators connected by an edge tend to settle into opposite phases (phase difference $\pi$)
    \item Oscillators not connected by an edge tend to synchronize in-phase
\end{itemize} 
The terminal configuration of the oscillator network directly equates to a resolution of the max-cut problem: 
\begin{itemize}
\item Oscillators with $0$ phase represent vertices in one set
   \item Oscillators with $\pi$ phase represent vertices in the other set
\end{itemize} 

The \textsc{Ising} Hamiltonian employed to describe the max-cut problem can be directly correlated to the energy of the coupled oscillator system, whereby minimizing the Hamiltonian is tantamount to maximizing the size of the cut. 

Anti-phase synchronization can be realized within oscillator networks without requiring an external second harmonic injection signal (see section  \ref{sec_implementation}), streamlining the hardware implementation for addressing max-cut problems. This intrinsic similarity between max-cut problems and anti-phase synchronization models facilitates an efficient hardware implementation of oscillator based \textsc{Ising} machines.
\section{Problem Mapping}
\begin{figure}
    \centering
    \includegraphics[width=0.15\textwidth]{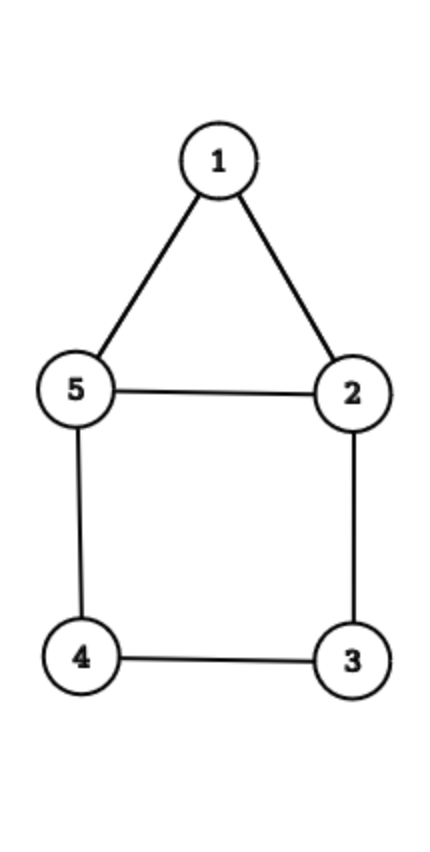}
    \caption{Typical max-cut graph}
    \label{pic_hvn}
\end{figure}

The mapping is therefore straightforward and corresponds to the adjacency matrix of the given graph. Initial computational tests were conducted using graphs with uniform weights, requiring adjustment of the resistive coupling between oscillators to set the system's initial conditions according to the problem graph. If the edge weight between two vertices in graph $G$ is $\mu_{ij}$, then as per \eqref{eq:ising_model_equation}, $J=-\mu_{ij}$. Thus, the resultant adjacency matrix indicates which oscillators are to be connected. The evolution of the system is then influenced by an external second harmonic injection lock (SHIL) signal to achieve binarization of the phase relationships of the oscillators. Details regarding the phase dynamics in these synchronizations and the resulting binarization are detailed in (\cite{wang2019oim}). Figure \ref{pic_hvn} shows a typical graph structure which is represented by the following (symmetric) coupling matrix:
\[
 C=\begin{pmatrix}
  0&1&0&0&1\\
  1&0&1&0&1\\
  0&1&0&1&0\\
  0&0&1&0&1\\
  1&1&0&1&0\\
 \end{pmatrix}
\]
\section{Implementation}
\label{sec_implementation}
The first design decision in building an \textsc{Ising} machine is the type 
and implementation of oscillators to be used. The main requirements here are 
simplicity and a $Q$ factor not too high to facilitate synchronization. 
Typical oscillator topologies could be ring oscillators, LC oscillators, or phase-shift oscillators. A discrete implementation as shown below makes 
ring-oscillators more involved than LC- or phase-shift-oscillators. Eventually
it was decided to implement phase-shift-oscillators due to their simplicity 
and the fact that they do not require physically large inductances. 

Another design decision is with respect to the synchronization technique: An
oscillator could either have a distince synchronization input and an output
or the output could also be used to feed a synchronization signal back into 
the oscillator. While the latter approach would half the number of connections
between oscillators in the case of symmetric couplings (the majority of 
problems can be formulated in such a way), the former approach seemed to be 
more suitable for a research system as all kinds of topologies could be 
configured, albeit at the cost of twice the number of connections for a 
symmetric problem.
\begin{figure}
 \centering
 \includegraphics[width=.8\textwidth]{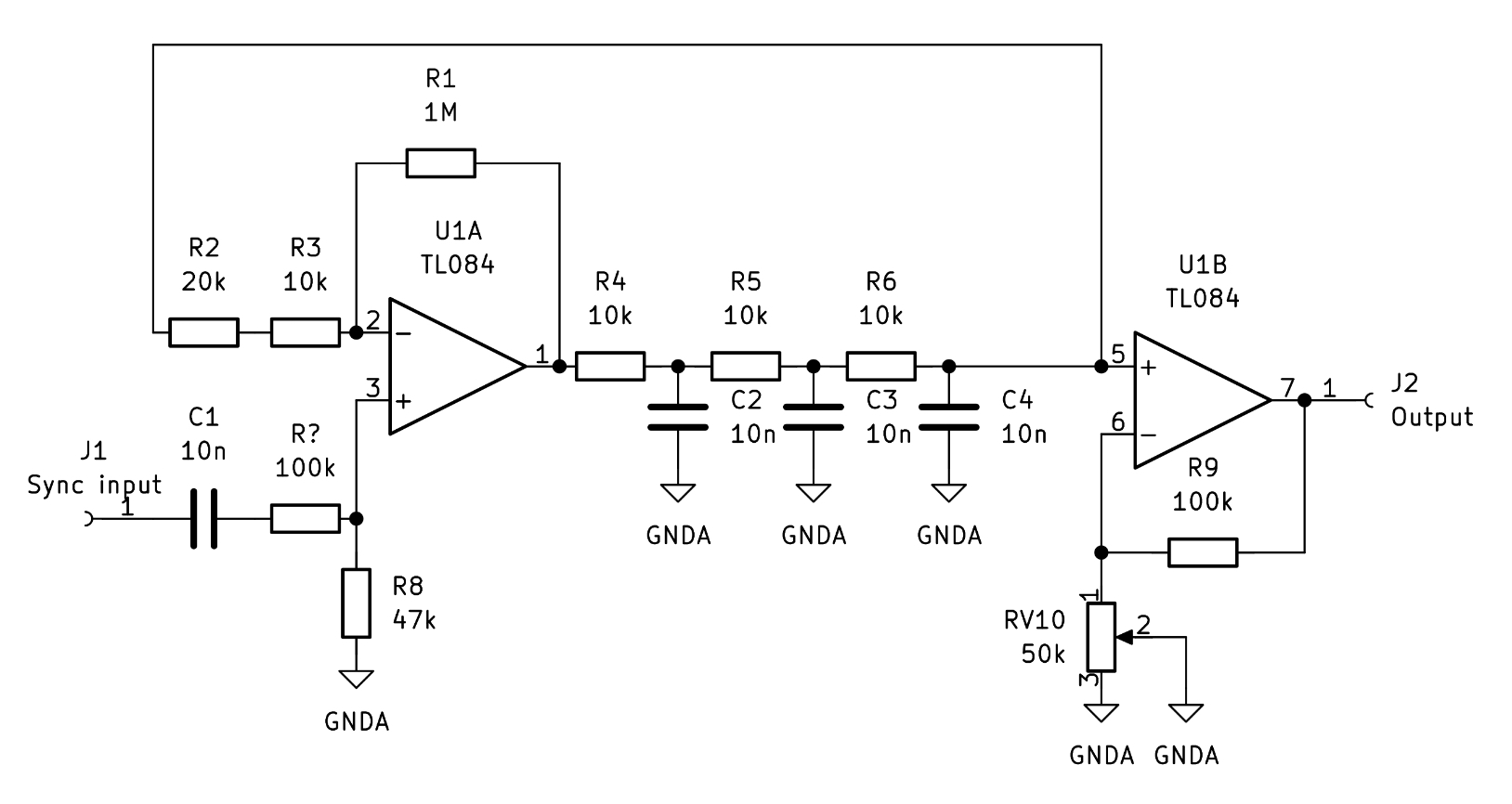}
 \caption{Schematic of the phase-shift-oscillator used}
 \label{pic_pso_schematic}
\end{figure}

Figure \ref{pic_pso_schematic} shows the schematic of the 
phase-shift-oscillators used in the following. At its heart is the operational
amplifier U1A followed by three frequency determining RC combinations 
R4/C2, R5/C2, and R6/C4. The output of this filter is coupled back into 
U1A and also fed into an output buffer amplifier U1B yielding a spectrally
quite clean output signal. The output amplitudes
of all oscillators is adjusted by RV10 to $\approx 4~\text{V}_\text{pp}$ 
with the oscillator running from a symmetric $\pm15$~V power supply. With 
the values specified in the schematic, the resonance frequency of these 
oscillators is about $3.8$~kHz. Since the remaining parts of the system are
comprised of off-the-shelf analog computer elements with a bandwidth of 
several $10$~kHz, even higher frequencies would be possible but not deemed
necessary for a proof-of-concept system. It should be noted that an 
oscillator's output signal will always be out of phase by $\pi$ with respect
to its synchronization input.

\begin{figure}
 \centering
 \includegraphics[width=\textwidth]{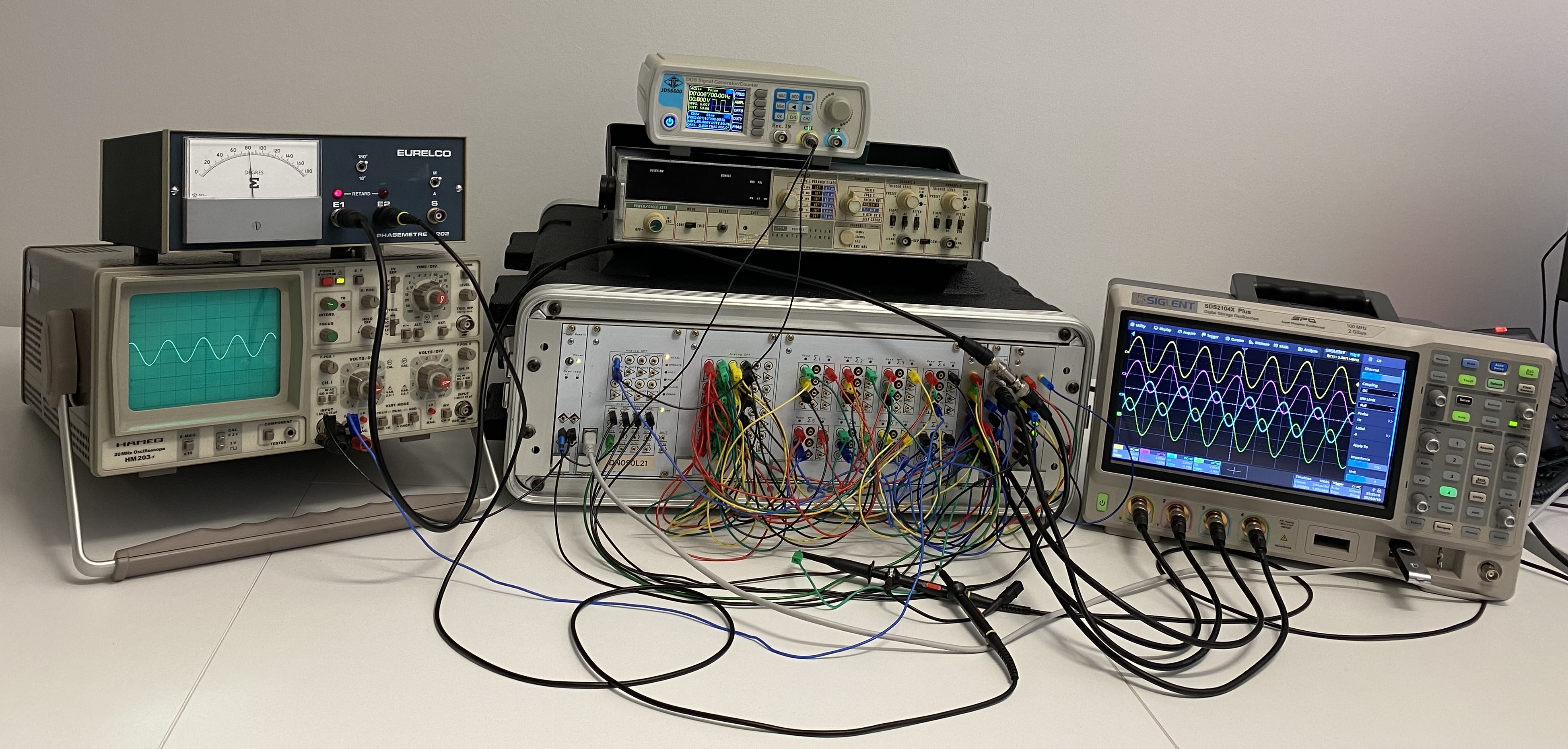}
 \caption{First incarnation of the oscillator based \textsc{Ising} machine}
 \label{pic_four_oscillators}
\end{figure}

The system described in the following started small with just four such
oscillators. This first implementation is shown in figure 
\ref{pic_four_oscillators}. The actual \textsc{Ising} machine is in the 
middle of the picture. On top of it is an external signal generator (see 
below) and a frequency counter. To its right is a four channel digital
oscilloscope displaying the output signals of all four oscillators while 
the oscilloscope on the left is used for debugging purposes. On top of this
oscilloscope is an analog phase meter that was initially used to determine
the phase relationships of the four oscillators.

This small machine allowed first exploratory experiments on graphs with 
up to four nodes. The oscillators can be coupled in an all-to-all fashion
using a matrix of digital potentiometers with $10$ bits of resolution.
Since coupling an oscillator with itself is not a useful connection, in
general only $n\times n-n$ coupling weights are required, $12$ in this 
case. 

To ensure that the oscillators can only lock into one of
two possible phases, $0$ and $\pi$, an external \emph{second harmonic 
injection locking} (\emph{SHIL}) signal is fed into all oscillators. This
forces the phases to \emph{binarize}. Thus, each oscillator's synchronization
input gets a sum of up to four signals: Three output signals from the other oscillators can be coupled with, and the SHIL signal. Synchronization
inputs are controlled by an analog switch, allowing an oscillator to run free or be synchronized.

\begin{figure}
 \centering
 \includegraphics[width=.6\textwidth]{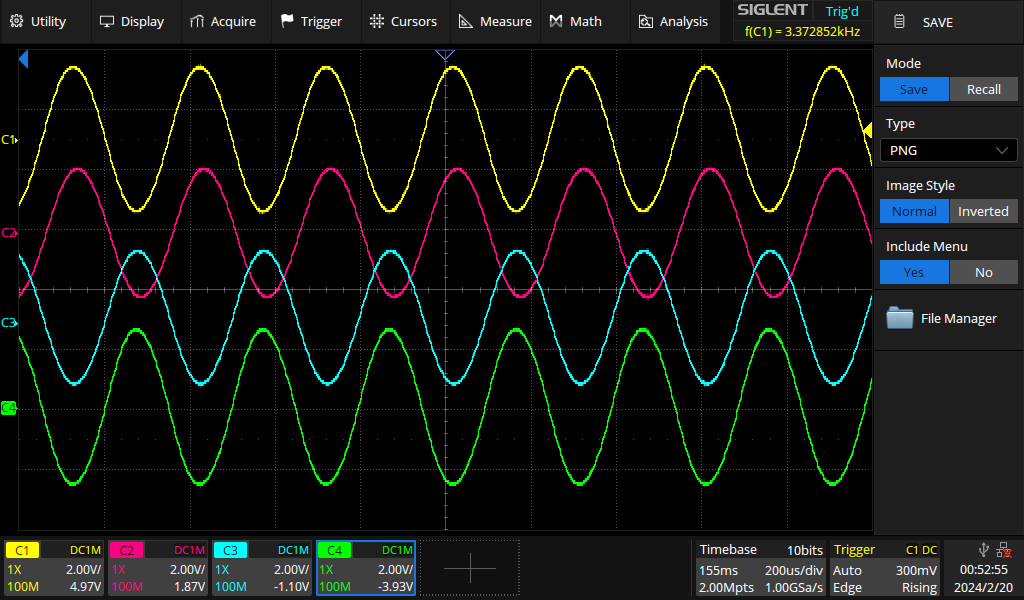}
 \caption{Typical solution on the four oscillator machine}
 \label{pic_typical_solution}
\end{figure}

Performing an actual computation requires the following steps: 
\begin{enumerate}
 \item All synchronization inputs are deactivated, so that the oscillators are free running (their phases will quickly randomize due to the very desirable and unavoidable differences in their resonance frequencies).
 \item The desired coupling weights are set by means of the digital 
  potentiometers (the actual values depend on the problem being solved, see
  section \ref{sec_coupling_weights}).
 \item Now all synchronization inputs are activated at once. It typically
  takes only a few oscillation periods for the oscillators to settle down 
  to a particular solution.
 \item Finally, the phases and thus the solution are read out by means of 
  some phase detector circuits. Figure \ref{pic_typical_solution} shows a 
  typical result for a certain connection graph. The actual solution, 
  \texttt{1100}, is encoded in the phase relationships of the oscillators 
  with respect to the first oscillator. 
\end{enumerate}

\begin{figure}
 \centering
 \includegraphics[width=.6\textwidth]{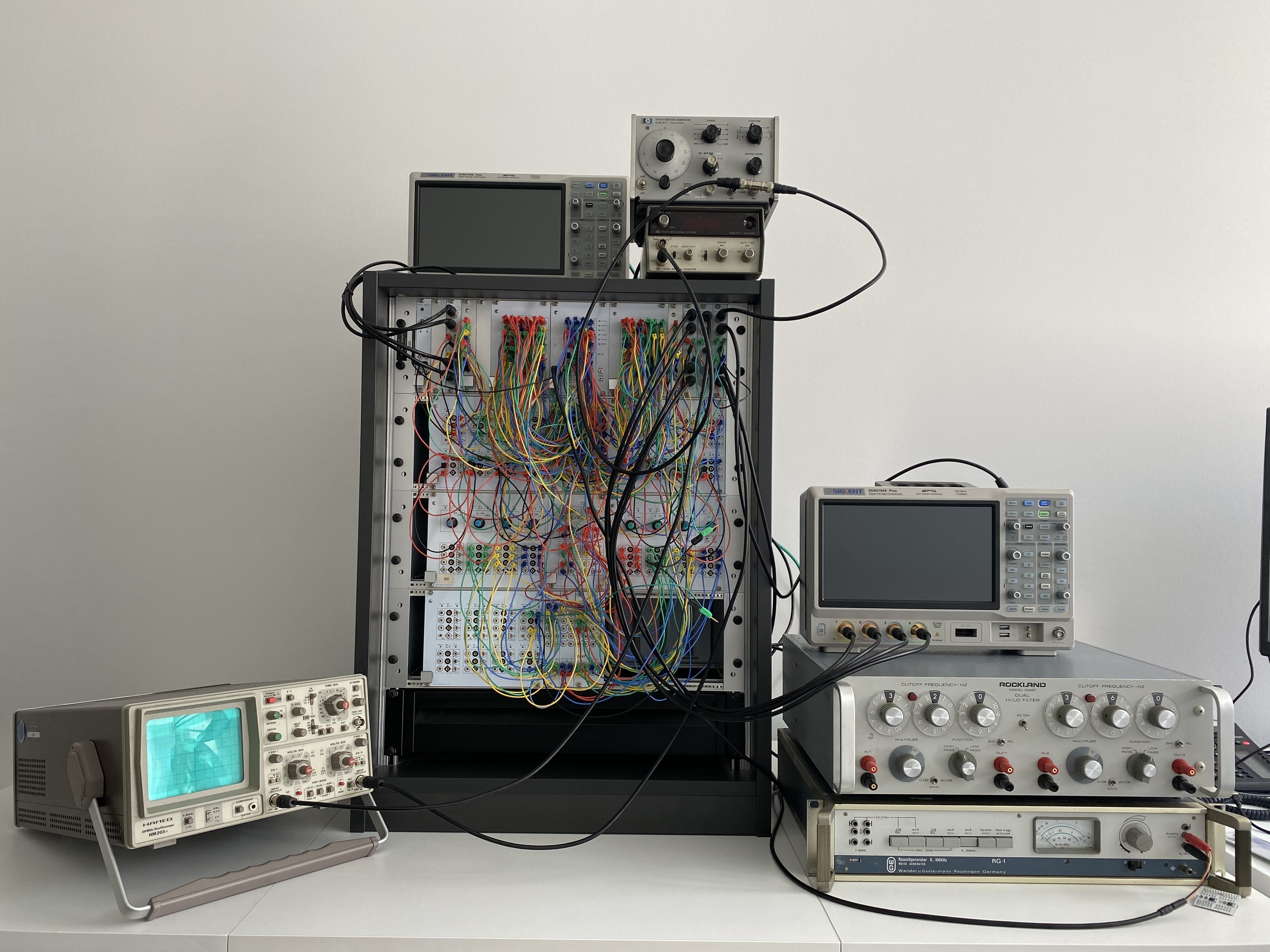}
 \caption{Eight oscillator \textsc{Ising} machine}
 \label{pic_8_oscillator_machine}
\end{figure}

\begin{figure}
 \centering
 \includegraphics[width=\textwidth]{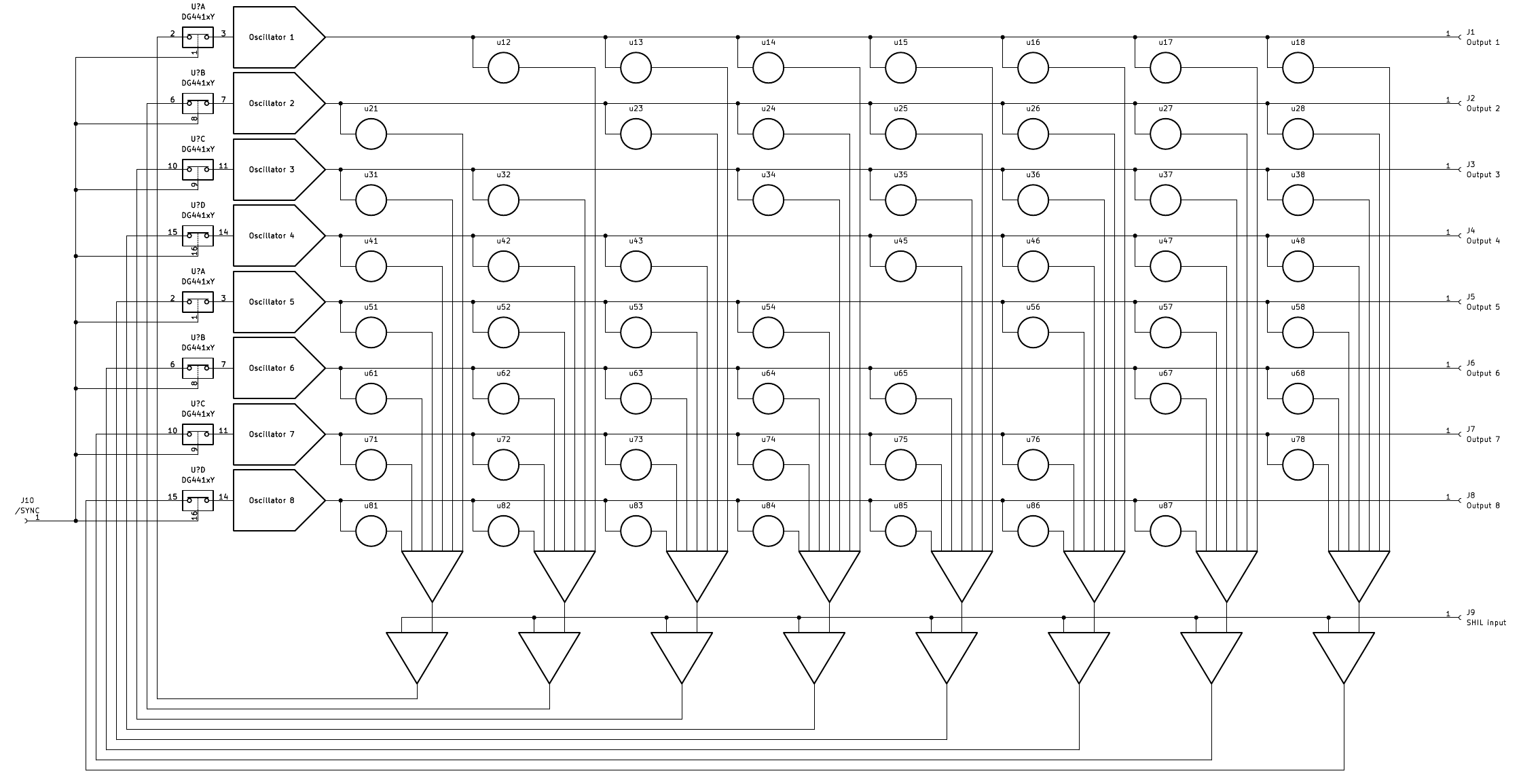}
 \caption{Schematic of the eight oscillator \textsc{Ising} machine}
 \label{pic_8_schematic}
\end{figure}

Due to the very good results obtained from experiments on this small machine, it was decided to expand it to eight oscillators, as shown in figure \ref{pic_8_oscillator_machine}. The structure of this machine is shown in figure \ref{pic_8_schematic}. At its heart are eight phase-shift oscillators of the type described above, feeding the rows of a $8\times 8$ matrix with zero diagonal, consisting of $56$ digital potentiometers with $10$ bits of resolution each. Each matrix column feeds a summer (with an implied change of sign due to its implementation), the output of which is fed into a second summer (thus reverting the first sign reversal) adding the external SHIL signal. The outputs of these eight summers are then fed to the respective oscillator's synchronization input by means of analog switches which are controlled externally.

This system, too, was built from standard analog computer modules from an anabrid GmbH \emph{Model-1} analog computer\footnote{See \url{https://analogparadigm.com/downloads/handbook.pdf}, retrieved 23.01.2025.} except for the oscillators which were purpose built. Using a simple hybrid controller\footnote{See \url{https://analogparadigm.com/downloads/hc_handbook.pdf}, retrieved 23.01.2025.} the system was coupled to a digital computer allowing it to control the digital potentiometers, the synchronization switches, and reading out the phases of the oscillator.

\begin{figure}
 \centering
 \includegraphics[width=.7\textwidth]{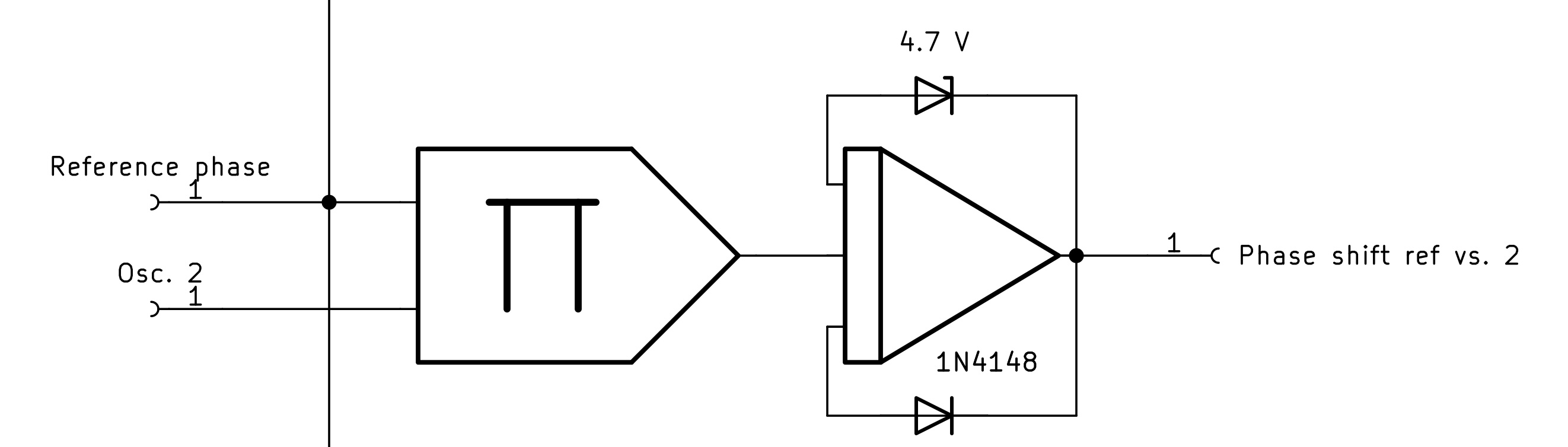}
 \caption{Schematic of one out of seven phase detectors used in the eight oscillator \textsc{Ising} machine}
 \label{pic_phase_detector}
\end{figure}

Since the phases typically binarize really well using an external SHIL signal, the phase detectors can be rather simple. Figure \ref{pic_phase_detector} shows the schematic of one out of seven phase detectors used in the machine. It consists of a multiplier that computes the product of the output of the reference oscillator and the oscillator under consideration for the measurement. The result of this multiplication will be $\sin^2(\omega t)$ if the oscillators are in phase and $-\sin^2(\omega t)$ if they are out of phase. This signal is then fed into an integrator having two additional diodes in its feedback loop. The 1N4148 diode at the bottom makes sure that the output signal cannot be negative\footnote{Due to the forward voltage of the diode, it can become slightly negative but that is not a problem here.}, the $4.7$~V \textsc{Zener} diode at the top clamps the output signal at $+4.7$~V maximum. If the phase of the oscillator is in phase with the reference oscillator, the output of the integrator will be $\approx 0$ while it will be $>0$ otherwise.

The outputs of seven of these simple phase detectors were connected to digital inputs of the hybrid controller used in the setup, so that the attached digital computer could read the phases after enabling the synchronization inputs of the oscillators and waiting for a few periods of oscillation. 
\section{Experiments and results}
Raw data and results from the experiments performed on the eight oscillator \textsc{Ising} machine will be provided by the authors upon request. These results are based on max-cut problems designed by the authors to serve as preliminary tests in order to determine if the machine yields correct solutions. Tests based on the $G$-set (a common benchmarking database for max-cut solvers, see \cite{stanford}) are yet to be conducted. 

Since only max-cut problems were tested, the phase relation ship of the oscillators representing nodes in the corresponding graphs directly correspond to the partition a particular node belongs to. If the graph contains seven nodes $1,\dots,7$ and the systems yields a solution \texttt{0101101}, this means that the max-cut partition of the graph is $\left\{1,3,6\right\}$ and $\left\{2,4,6,7\right\}$.

The analysis of results obtained by our machine show that different problems require varying coupling strengths to enhance the likelihood of producing the correct output. Even if all coupling weights are equal values $c$ (which is the case for all problems considered until now), the actual value $c$ has to be chosen individually for every problem to achieve best results, although values like $c=\frac{1}{10}$ or $c=\frac{1}{5}$ tend to be good candidates for most problem graphs. This is most likely caused by the following potential contributing factors:
\begin{description}
    \item [\textbf{Node Connectivity}:] Oscillators with a higher number of connections tend exhibit anomalous behavior, potentially due to ``frustration'' caused by excessive interactions.
    \item [\textbf{Closed odd loops}:] Graph structures containing closed odd-numbered loops (e.\,g. triangles, pentagons, heptagons, etc.) display unique synchronization behaviors, often partitioning into different configurations upon repeated runs (\cite{belykh2005synchronization}), which is not unexpected given that only two phase relationships can be reached which clashes with loops consisting of an odd number of nodes.
    \item [\textbf{Graph depth}:] The depth of the graph also plays a role in determining the machine’s performance, though its precise influence requires further investigation (\cite{belykh2005synchronization}).
    \item [\textbf{Effect of SHIL}:] The experiments were carried out with and without the presence of a second harmonic injection line (SHIL) at 2.5 V and 6.8 kHz, using sinusiodal, rectangular, and pulse signals. This configuration consistently produced accurate solutions and represents the ``Goldilocks zone'' for most of the problems tested.
    \item [\textbf{The problem itself}:] Some problems do not require a SHIL signal while others do. The max-cut problem for $G(5,6)$ (commonly called the house graph) is an example for the latter. In the absence of SHIL, this particular system repeatedly produces incorrect solutions.
    \item [\textbf{Graph isomorphism}:] Analysis of isomorphic problems revealed consistently correct solutions under the influence of SHIL, regardless of oscillator assignments (as was to be expected, though). Examples include rectangle, toffee, crisscross12\_78bar, and hourglass graphs. These are all isomorphic to a ring graph with 8 vertices (\cite{belykh2005synchronization}).
    \item [\textbf{Synchronization dynamics}:] Closed odd-numbered loops induce synchronization into distinct partitions with each execution. 
\end{description}
\section{Open research questions}
There is a number of open questions resulting from the above work which are described in the following sections.
\subsection{Oscillator phase response curve}
The \emph{phase response curve} (\emph{PRC}) of the system remains to be determined. The PRC delineates how an oscillator's phase is altered in response to external stimuli, such as coupling with neighboring oscillators or external signals like the SHIL. Comprehending the PRC is paramount for anticipating synchronization behavior and stability in networks of coupled oscillators. In a further step, a thorough mathematical analysis of the PRC is required.
\subsection{Solutions} 
\emph{Twist case} solutions represent states where oscillators assume a distinct phase-difference pattern, often periodic or symmetric (\cite{bolotov2019twisted}). A better understanding of these states is imperative to recognize patterns and optimize system performance for specific computational tasks. 
\subsection{Phase dynamics, SHIL, etc.}
Examination of the phase dynamics requires an analysis of how the phases of coupled oscillators evolve over time when subjected to external synchronization signals. This involves a review of the phenomena of phase locking, synchronization, and desynchronization and their reliance on SHIL parameters.

Another question is whether it might be worthwhile to introduce noise from a noise generator into the system (perhaps a noise SHIL signal) to explore more of the possible state space of a given topology. This might offer a way to leave local minima unaffected by noise. It is to be determined what characteristics a suitable noise source must fulfill for this purpose (bandwidth, amplitude, probability distribution). \cite{vaidya} contains a lot of information on this question.

How should coupling weights (including the SHIL signal) be handled in general? The system described above activates all coupling weights of a problem set-up at once. It might be worthwhile to gradually increase the coupling weights instead. 

What about SHIL? What effect would a slowly increasing ramp-up of the injected signal have on the quality of the solutions obtained by the machine? Maybe weights could also be ``wiggled'' by means of a pseudo-random number generator under control of the attached digital computer in order to explore a larger problem space (instead of adding a noise signal to the SHIL).

What type of SHIL signal should be used in general? Is there a general rule for this? The experiments showed that a sinusoidal SHIL signal did not work as well as a rectangular one or a pulse signal on most setups. Is an external SHIL signal even required in all cases? 
\subsection{Binarization and beyond}
Another observation is that using a SHIL signal with three times the resonance frequency of the oscillators comprising the \textsc{Ising} machine the phase relationships of the oscillators will ternarize. This makes it possible to not just represent states $\left\{0,1\}\right\}$ but $\left\{-1,0,1\right\}$. Such a ternarized system might very well prove to be computationally even more powerful than a system with a SHIL of twice the resonance frequency of the oscillators.
\subsection{Suitable oscillator types}
Alternate oscillator types, such as LC circuits (inductor-capacitor) or even \emph{phase-locked loops} (\emph{PLL}s), confer distinct advantages in terms of frequency stability, coupling flexibility and energy efficiency. Investigating these implementations may improve system performance and scalability.
\subsection{Hardware topologies and coupling weights}
\label{sec_coupling_weights}
The most desirable interconnect structure of such an \textsc{Ising}-machine would be an all-to-all connection graph. Unfortunately this becomes unfeasible quickly as the number of weights required for $n\in\mathbb{N}$ oscillators it $n^2-n$. Consequently, techniques for the optimization of connectivity and problem mapping are required. Sparse but highly interconnected topologies might reduce energy consumption and enhance solution accuracy, while adaptable mapping mechanisms allow for the resolution of a wider array of problems.

Scaling the system to larger networks introduces additional challenges, such as increased coupling complexity, communication delays, and cross coupling. Investigating these effects is crucial for preserving the quality of the solution and the efficiency of the hardware as the system expands.

Another open question is how many bits of resolution are required per coupling weight. The experiments described above all used coupling weights with $10$ bits of resolution of which only a very small fraction were actually used or required. What is the lower limit for coupling weights? From an implementation perspective, single bit weights would be ideal as this would vastly simplify the interconnect circuitry. Also, ternary weights $\left\{-1,0,1\right\}$ would be rather easy to implement. Would this suffice for a wide range of problem classes?
\subsection{Going large scale}
Realizing a large-scale \textsc{Ising} machine with at least $10^4$ oscillators requires advancements in hardware integration, interconnect structure, coupling weight implementation, and synchronization control. Such a system could effectively address complex combinatorial optimization problems, heralding real-world applications in domains such as AI, logistics, bio-informatics and many more. 

Identifying the coupling function of the described system is equally important. The coupling function regulates how oscillators interact, affecting their phases, frequencies, and synchronization dynamics. Various forms of coupling (e.\,g., linear, nonlinear, or time-varying) can optimize performance for particular topologies and problem types, with coupling strength greatly influencing stability and precision. Efficient coupling mechanisms must balance energy dissipation, scalability, and robustness against external influences such as SHIL. Customizing coupling functions for specific applications, oscillator types (e.g., LC circuits or PLLs), and hardware constraints enables enhanced synchronization and computational efficiency. The empirical validation of coupling models and their adaptation for large-scale systems are vital to ensuring reliable performance as networks expand to encompass thousands of oscillators.
\conclusions  
It may be concluded that \textsc{Ising} machines show great promise as a computational paradigm when it comes to computationally complex problems (NP hard), although a number of research and engineering questions are still to be answered. 

The main engineering challenges are scaling such systems up to $10^n$ oscillators with $n>>4$, ideally $n>>6$. Devising and implementing suitable interconnect structures is a challenge as is avoiding spurious cross talk between the oscillators.

In addition to this a number of research questions remains unanswered as of now.
%
\codedataavailability{Code and data sets may be requested from the authors.} 
\authorcontribution{The authors contributed equally to the work described in this publication.} 

\competinginterests{There are no competing interests.} 

\begin{acknowledgements}
 The authors would like to thank Dr. \textsc{Lucas Wetzel} for many in-depth
 discussions and suggestions regarding our work on oscillator based 
 \textsc{Ising} machines.
\end{acknowledgements}

\end{document}